\documentclass[conference,10pt]{IEEEtran}
\usepackage{epsfig,epsf,rotating,setspace,latexsym,amsmath,amssymb,amsfonts,bm,theorem,subfigure,epstopdf}
\usepackage{cite,authblk}
\usepackage{bbm}
\usepackage{algorithm}
\usepackage[noend]{algpseudocode}
\usepackage{color}
\usepackage{mathtools}
\usepackage{soul}

\algrenewcommand\algorithmicforall{\textbf{foreach}}
\algrenewcommand\algorithmicindent{.8em}

\newtheorem{lemma}{Lemma}

\newenvironment{Proof}[1]{\medskip\par\noindent{\bf Proof:\,}\,#1}{{\mbox{\,$\blacksquare$}\par}}

\IEEEoverridecommandlockouts
\allowdisplaybreaks

\begin{document}
 
\title{Freshness Based Cache Updating in \\ Parallel Relay Networks}
 
\author{Priyanka Kaswan \qquad Melih Bastopcu \qquad Sennur Ulukus\\
        \normalsize Department of Electrical and Computer Engineering\\
        \normalsize University of Maryland, College Park, MD 20742\\
        \normalsize  \emph{pkaswan@umd.edu} \qquad \emph{bastopcu@umd.edu} \qquad \emph{ulukus@umd.edu}}
 
\maketitle

\begin{abstract}
We consider a system consisting of a server, which receives updates for $N$ files according to independent Poisson processes. The goal of the server is to deliver the latest version of the files to the user through a parallel network of $K$ caches. We consider an update received by the user successful, if the user receives the same file version that is currently prevailing at the server. We derive an analytical expression for information freshness at the user. We observe that freshness for a file increases with increase in consolidation of rates across caches. To solve the multi-cache problem, we first solve the auxiliary problem of a single-cache system. We then rework this auxiliary solution to our parallel-cache network by consolidating rates to single routes as much as possible. This yields an approximate (sub-optimal) solution for the original problem. We provide an upper bound on the gap between the sub-optimal solution and the optimal solution. Numerical results show that the sub-optimal policy closely approximates the optimal policy.
\end{abstract}

\section{Introduction}
In the information age, users want instant access to up-to-date data. Caching is a popular method of pre-storing data at nodes in a network closer to the users for faster delivery of latest data. In recent years, various papers have explored freshness-optimal policies in different settings. Most works have relied on the age of information (AoI) metric to measure obsoleteness of data. AoI has been considered in a wide range of contexts, such as queueing networks, energy harvesting systems, web crawling, scheduling problems, remote estimation, UAV systems and so on \cite{Najm17,Soysal19,cho03,kolobov19,Farazi18,Wu18,Baknina18,Leng19,Arafa20,Elmagid20,Ceran18,liu18,elmagidUAV19,Bastopcu20_soft_updates,Bastopcu20_group,yates19_status_update,kadota18,hsu18,Buyukates19_hier,Bedewy19,Buyukates19_multihop,wang19_counting,bastopcu20_google,sun17_remote,Bastopcu20_infection,yun18,kam20,chakravorty20,mayekar20,Bastopcu20_selective,ramirez20,Buyukates20_stragglers,zou19,ozfatura20,yang20,Rajaraman18,Bastopcu19_distortion,Ayan19,Banerjee20,Yates17sqrt,Tang19FileAge,bastopcu2020LineNetwork,bastopcu2020LimitedCache,Zhang18MobileEdgeCaching,Gao12opportunistic,Pappas20EH,Kam17HistoryFreshness,gu2020TwoHop,Arafa19TwoHop}. 

The works that are most closely related to our work here are \cite{Yates17sqrt,Tang19FileAge,Zhang18MobileEdgeCaching,Gao12opportunistic,Pappas20EH,Kam17HistoryFreshness,gu2020TwoHop,Arafa19TwoHop,bastopcu2020LimitedCache,bastopcu2020LineNetwork}. In \cite{Yates17sqrt}, a single-server single-cache refresh system is considered, where  it  is  shown  that  an  asymptotically  optimal  policy updates  a  cached  file  in  proportion  to  the  square  root  of its popularity. The work in \cite{Yates17sqrt} assumes constant file update durations, which is extended in \cite{Tang19FileAge} by considering file update durations to be dependent on the size and the age of the files. While \cite{Yates17sqrt,Tang19FileAge} use the AoI metric, reference \cite{bastopcu2020LineNetwork} uses a binary freshness metric in a caching system, and determines the optimum update rates at the user and the cache. \cite{bastopcu2020LineNetwork} also extends the approach to a cascade sequence of cache nodes, and  \cite{bastopcu2020LimitedCache} generalizes it to the case of nodes with limited cache capacity. Here, we further generalize \cite{bastopcu2020LineNetwork} to a more complex network which is composed of parallel caches.  

Other related work that use caching and relaying techniques for freshness include: \cite{Zhang18MobileEdgeCaching} where a tradeoff between content freshness and service latency from the aspect of mobile edge caching is studied; \cite{Gao12opportunistic} which considers caching policies in opportunistic networks;  
\cite{Pappas20EH} where a cache-enabled aggregator decides whether to receive a fresh update from an energy harvesting sensor or serve the request with a cached update; \cite{Kam17HistoryFreshness} where an optimal policy is derived when the current rate of requests for a file is dependent on both history of requests and the freshness of the file; \cite{gu2020TwoHop} which considers a two-hop status update system where an optimal scheduling policy is identified by a constrained Markov decision process approach; and \cite{Arafa19TwoHop} where a two-hop system with energy harvesting at source and relay nodes is considered. 

\begin{figure}[t]
\centerline{\includegraphics[width=0.85\linewidth]{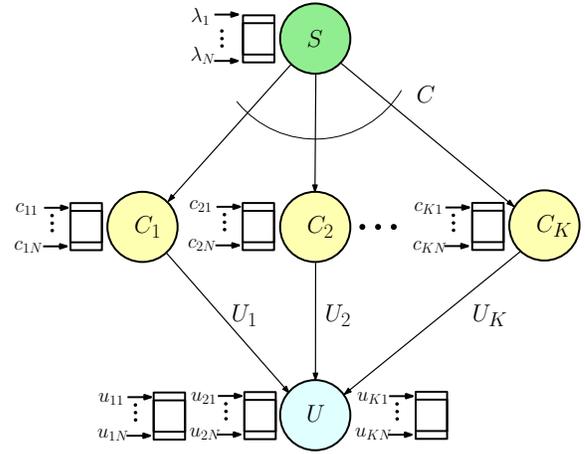}}
\caption{System model for a parallel multi-cache system.}
\label{fig:system_model}
\vspace*{-0.4cm}
\end{figure}

In this paper, we consider a parallel network with multiple cache routes between a source and a user (Fig.~\ref{fig:system_model}). We first derive a closed-form expression for freshness at the user. We observe from the freshness formula of the two-cache system that lop-sided distribution of rates across the routes supports higher freshness. Further, for the two-route two-file case, restricting at least one of the files to a single route maximizes the overall freshness of the system. Moreover, in a $K$-cache system, restricting a file to fewer routes improves the freshness. Motivated by these properties, we solve an auxiliary problem of a single-cache system and adapt its solution to our parallel-cache network to obtain an approximate (sub-optimal) solution for the original problem. We provide an upper bound on the gap between the sub-optimal policy and the optimal policy. The gap is finite and is independent of the number of files. Numerical results show that the proposed sub-optimal policy closely approximates the optimal policy. 

\section{System Model and Problem Formulation} \label{sect:system-model}
We consider a system with a source, $K$ parallel relays and a user, as shown in Fig.~\ref{fig:system_model}. The source has the most up-to-date versions of a library of $N$ files. File $i$ is updated at the source with exponential inter-update times with rate $\lambda_i$. The source updates file $i$ at cache $k$ with exponential inter-update times with rate $c_{ki}$. Cache $k$ updates file $i$ at the user with exponential inter-update times with rate $u_{ki}$. There is no delay or information loss in any source-cache links or cache-user links. However, the source is subject to a total update rate constraint $\sum_{k=1}^{K}\sum_{i=1}^{N}c_{ki}\leq C$, and cache $k$ is subject to a total update rate constraint $\sum_{i=1}^{N}u_{ki}\leq U_{k}$, for $k=1,\ldots,K$. 

When a file is updated at the source, the stored versions of the same file at the caches and at the user become outdated. Thus, we consider an update received by the user successful if the user receives a file version that is currently prevailing at the source. This will happen when the source updates the cache and the cache in turn updates the user before the file at the source is updated with a newer version. In the following subsections, we first derive a freshness expression for file $i$ in a single-cache system, and then in a multi-cache system. For simplicity, we drop subscript $i$ from $\lambda_i$, $c_{ki}$ and $u_{ki}$ since the derivation is valid for all files (for all $i$).

\subsection{Freshness of File $i$ in the Single-Cache Model}
\begin{figure}[t]
\begin{center}
\subfigure[]{%
\includegraphics[width=0.49\linewidth]{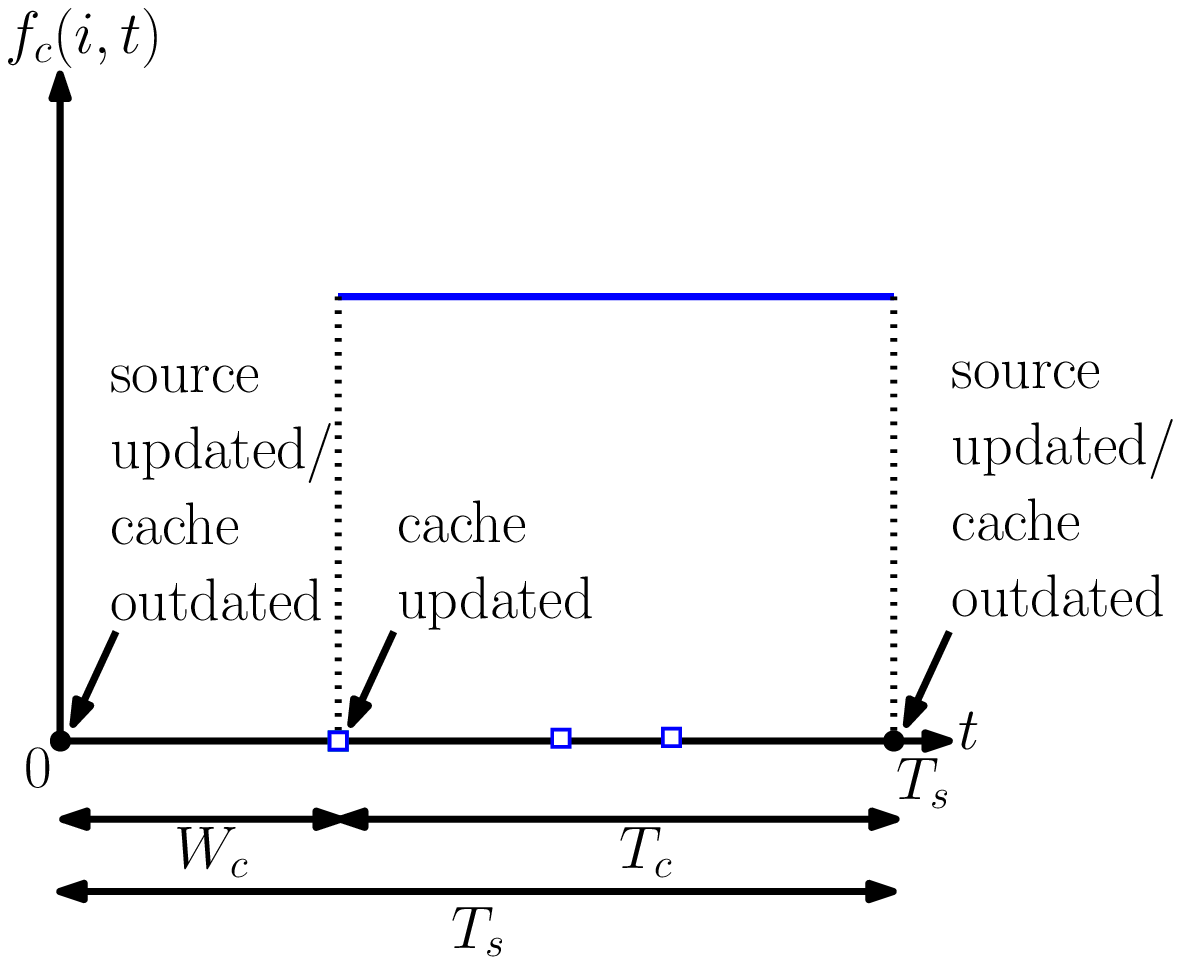}}
\subfigure[]{%
\includegraphics[width=0.49\linewidth]{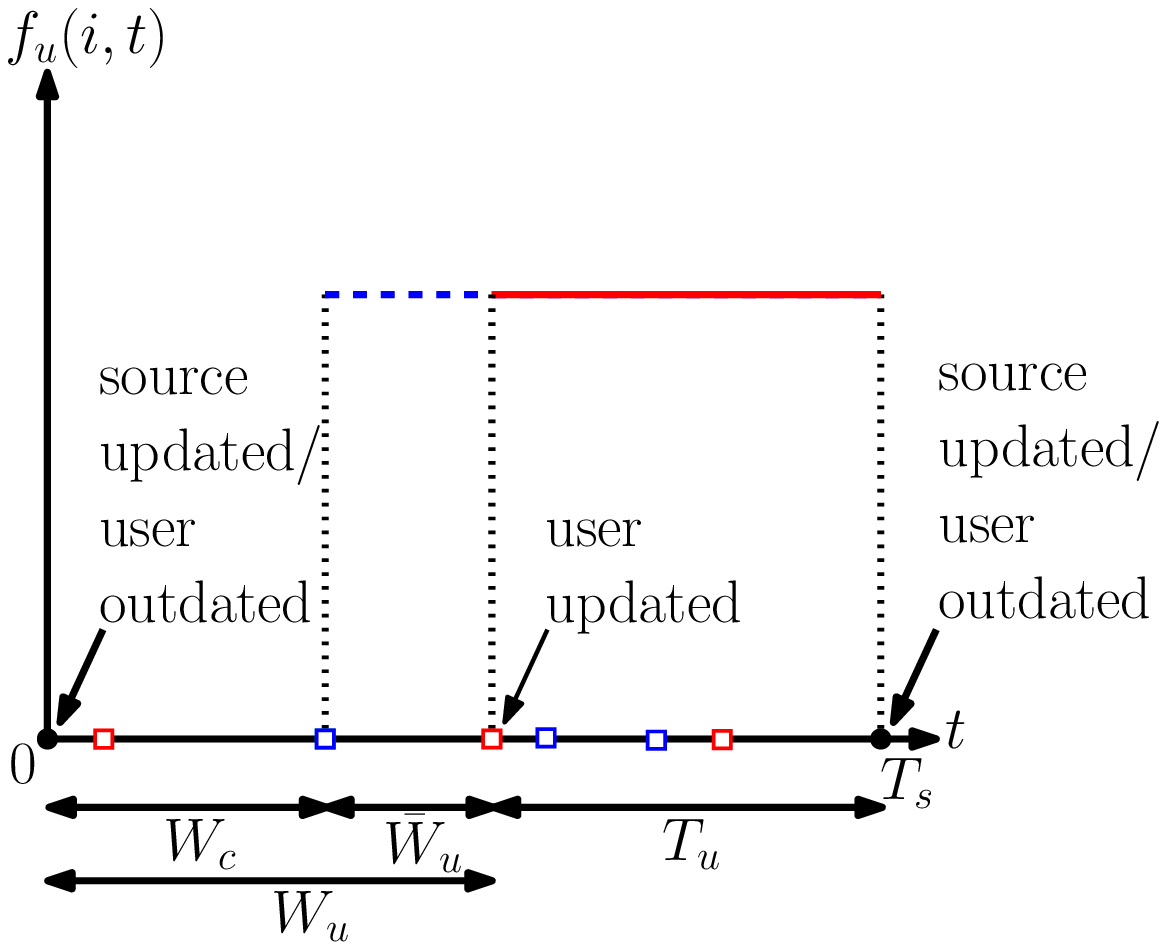}}
\end{center}
\vspace{-0.5cm}
\caption{Freshness function as a function of time at the cache and the user.}
\label{fig:var_explain}
\vspace{-0.4cm}
\end{figure}

In this subsection, we find the freshness expression for file $i$ for a single-cache system. First, we characterize the freshness at the cache. In Fig.~\ref{fig:var_explain}(a), the freshness evolution at the cache is shown between two file updates at the source. We define the freshness function for file $i$ at the cache as follows
\begin{align}
f_{c}(i,t) = \begin{cases} 
1, & \text{if file $i$ at the cache is fresh at time $t$},\\
0, & \text{otherwise.}
\end{cases}
\end{align}

Let $T_{s}(i,j)$ denote the $j$th update cycle at the source, i.e., time interval between the $j$th and $(j+1)$th update for file $i$. Once the source gets updated, the cache is updated after duration $W_{c}(i,j)$ and it remains updated for $T_{c}(i,j)=T_s(i,j)-W_c(i,j)$ duration. For simplicity, we drop index $i$ for variables $T_{s}(i,j)$, $T_{c}(i,j)$, and $W_{c}(i,j)$, as the results in this subsection pertain to file $i$. We denote $F_{c}(i)$ as the long term average freshness of file $i$ at the cache which is given by
\begin{align}\label{F_c(i)_temp}
F_{c}(i) = \lim_{T\rightarrow\infty} \frac{1}{T} \int_{0}^{T} f_{c}(i,t)dt. 
\end{align}
Let $M$ be the number of update cycles in time duration $T$. Provided that the system is ergodic, similar to \cite{bastopcu2020LineNetwork}, $F_c(i)$ is 
\begin{align}\label{F_c(i)}
F_{c}(i) = \lim\limits_{T\rightarrow \infty}\frac{M}{T}\left(\frac{1}{M}\sum_{j=1}^{M}T_{c}(j)\right)
= \frac{\mathbb{E}[T_{c}]}{\mathbb{E}[T_{s}]}.
\end{align}
Here, as $T_c(j)$ are independent and identically distributed (i.i.d.) over $j$, we drop the index $j$ and denote $T_c(j)$ with the typical random variable $T_c$. Similarly, $T_s$ and $W_c$ denote the typical random variables for $T_s(j)$ and $W_c(j)$, respectively.

Since $T_s$ is an exponential random variable with rate $\lambda$, we have  $\mathbb{E}[T_s]=\frac{1}{\lambda}$. We find $\mathbb{E}[W_c]$ by using nested expectations, i.e., $\mathbb{E}[W_c] = \mathbb{E}[\mathbb{E}[W_c|T_s]]$. For a given update cycle duration $T_s=t$ at the source, $W_c$, which is exponentially distributed with rate $c$, either takes a value in between $0$ and $T_s$, or takes value $T_s$, i.e., $W_c=T_s$, and in this case, the cache is not updated in that cycle. Thus, we have
\begin{align} \label{E[W_c/T_s=t]}
\!\!\mathbb{E}[W_c|T_s\!=\!t]\!=\!\int_{0}^{t}\!\!xce^{-cx}dx + \int_{t}^{\infty}\!\!\!\!tce^{-cx}dx =\frac{1-e^{-ct}}{c}.
\end{align}
Then, we obtain $\mathbb{E}[W_c]$ as
\begin{align} \label{E[W_c]}
\mathbb{E}[W_c]= \int_{0}^{\infty}\frac{1-e^{-ct}}{c} \lambda e^{-\lambda t}dt
=\frac{1}{\lambda+c}.
\end{align}
 By using (\ref{E[W_c]}), we obtain $\mathbb{E}[T_c]=\mathbb{E}[T_s]-\mathbb{E}[W_c]$ as
\begin{align} \label{E[T_c]}
\mathbb{E}[T_c]=\frac{c}{\lambda(\lambda+c)}.
\end{align}
Finally, by substituting (\ref{E[T_c]}) into (\ref{F_c(i)}), we obtain $F_c(i)$ as
\begin{align}
\label{F_c}
F_c(i)=\frac{c}{\lambda+c}.
\end{align}

Next, we characterize the freshness at the user. Freshness evolution at user in an update cycle is shown in Fig.~\ref{fig:var_explain}(b). We define the freshness function for file $i$ at the user as follows
\begin{align}
f_{u}(i,t) = \begin{cases} 
1, & \text{if file $i$ at the user is fresh at time $t$}, \\
0, & \text{otherwise.}
\end{cases}
\end{align}
Once file $i$ is updated at the cache after $W_c(j)$, and the same file is updated at the user after $\bar{W}_u(j)$, file $i$ at the user remains fresh for a time period of $T_u(j)$. Thus, the total waiting time for the user to get the freshest version of file $i$ in the $j$th cycle is $W_u= W_c+\bar{W}_u$. We denote $F_{u}(i)$ as the long term average freshness of the file $i$ at the user which is given by 
\begin{align} \label{F_u(i)}
F_{u}(i)  = \frac{\mathbb{E}[T_{u}]}{\mathbb{E}[T_{s}]},
\end{align}
where $T_u$ denotes the typical random variable for $T_u(j)$.

First, we find $\mathbb{E}[W_u]$ by using nested expectations. When the $j$th update arrives at the source, due to memoryless property of the exponential distribution, $W_c$ and $\bar{W}_u$ are exponentially distributed with rates $c$ and $u$, respectively. Hence, the distribution of $W_u$ denoted by $f_{W_u}(x)$ is equal to the convolution of the distributions of $W_c$ and $\bar{W}_u$ which is given by 
\begin{align} \label{pdf_of_w_u}
f_{W_u}(x)=\frac{cu}{c-u}\left(e^{-ux}-e^{-cx}\right), \quad 0\leq x< \infty.
\end{align}
For a given update cycle duration $T_s=t$ at the source, the total waiting time $W_u$ with pdf in (\ref{pdf_of_w_u}) either takes a value in between $0$ and $T_s$, or $W_u=T_s$. When $W_u=T_s$, we note that the file at the user is not updated in that cycle. Thus, 
\begin{align}
\mathbb{E}[W_u|T_s=t]= \frac{cu}{c-u}\left(\frac{1-e^{-ut}}{u^2}-\frac{1-e^{-ct}}{c^2}\right) .
\end{align}
By using $\mathbb{E} [ W_u] =\mathbb{E}[\mathbb{E}[W_u|T_s]]$, we obtain $\mathbb{E} [ W_u]$ as
\begin{align} \label{E[W_u]}
\mathbb{E}[W_u]= \frac{\lambda+c+u}{(\lambda+u)(\lambda+c)}.
\end{align}
Then, we obtain $\mathbb{E}[T_u]=\mathbb{E}[T_s]-\mathbb{E}[W_u]$ as
\begin{align} \label{E[T_u]}
\mathbb{E}[T_u]=\frac{uc}{\lambda(\lambda+u)(\lambda+c)}.
\end{align}
Finally, by substituting  (\ref{E[T_u]}) into (\ref{F_u(i)}), we obtain $F_u(i)$ as
\begin{align}\label{F_u(i)final}
F_u(i)=\frac{\mathbb{E}[T_u]}{\mathbb{E}[T_s]}=\frac{u}{\lambda+u}\frac{c}{\lambda+c},
\end{align}
which is equal to the freshness expression in \cite{bastopcu2020LineNetwork}. Above, we have provided an alternative method (to \cite{bastopcu2020LineNetwork}) to derive freshness, which will be useful in the multi-cache system next. 

\subsection{Freshness of File $i$ in the Multi-Cache Model}
In this subsection, we find the freshness expression of file $i$ for a multi-cache system. For simplicity, we drop file index $i$ from all variables. Each cache sends its updates to the user independent of other caches. After the file at the source is updated for the $j$th time, the file at the user becomes fresh again by the first successful update by any one of the caches. The file at the cache $k$ is updated after $W_{c_k}$ duration. The cache $k$ updates the same file at the user after $\bar{W}_{u_k}$ duration. We denote the random variable $X_k=W_{c_k}+\bar{W}_{u_k}$ as the total waiting time for cache $k$ to send a successful update to the user. As $W_{c_k}$ and $\bar{W}_{u_k}$ are exponentially distributed with rates $c_k$ and $u_k$, respectively, similar to (\ref{pdf_of_w_u}), we have  $f_{X_k}(x)=\frac{c_ku_k}{c_k-u_k}\left(e^{-u_kx}-e^{-c_kx}\right)$ for $x\geq 0$. For a given update cycle $T_s=t$, the user is updated after $W_u$ given by
\begin{align}
\label{min_waiting_time}
W_u=\min\{t,X_1,X_2,\ldots,X_K\},
\end{align}
where $W_u =t$ denotes the case where the user is not updated in that update cycle. The ccdf of $X_k$ is given by
\begin{align}
\mathbb{P}(X_k>x)= \begin{cases} 
\frac{c_ku_k}{c_k-u_k}\left(\frac{e^{-u_k x}}{u_k}-\frac{e^{-c_k x}}{c_k}\right), & x\geq 0,\\
1,& x<0.
\end{cases}
\end{align}

Since $W_u$ takes only positive values, $\mathbb{E}[W_u]$ can be found by integrating its ccdf, i.e., $\mathbb{E}[W_u|T_s=t] = \int_{0}^{\infty}\mathbb{P}(W_u>x)dx$, 
\begin{align}
\label{conditional_min_waiting_time}
\mathbb{E}[W_u|T_s=t] \!=\!\int_{0}^{t} \mathbb{P}(X_1>x) \cdots \mathbb{P}(X_K>x)dx.
\end{align}
For ease of exposition, let $p_v=(p_i)_{i \in [k]} \in \Pi_{k }\{c_k, u_k \}=V_p$, and $S_c=\sum_{k=1}^{K} \mathbbm{1}{\{p_k=c_k\}}$. Then, $\mathbb{E}[W_u|T_s=t]$ equals
\begin{align}
\frac{\prod_{k}c_k\prod_{k}u_k}{\prod_{k}(c_k-u_k)} 
\left(\sum_{p_v \in V_p}\frac{(-1)^{S_c}\left(1-e^{-t\left(\sum_{k}p_k\right)}\right)}{\left(\sum_{k}p_k\right)\prod_{k}p_k}\right).
\end{align}

Next, we find $\mathbb{E}[W_u] = \mathbb{E}[\mathbb{E}[W_u|T_s]]$ as
\begin{align}
\mathbb{E}[W_u]=&\frac{\prod_{k}c_k\prod_{k}u_k}{\prod_{k}(c_k-u_k)}\left(\sum_{  p_v \in V_p}\frac{(-1)^{S_c}}{\prod_{k}p_k}\frac{1}{\lambda+\sum_{k}p_k}\right).
\end{align}
Since $\mathbb{E}[T_u] =\mathbb{E}[T_s]-\mathbb{E}[W_u]$, we find $F_u(i)$ in \eqref{F_u(i)} as
\begin{align} \label{F_u_i}
F_u(i) \!=\! 1\!-\!\frac{\prod_{k}c_k\prod_{k}u_k}{\prod_{k}(c_k-u_k)}\left(\!\sum_{ p_v  \in V_p}\frac{(-1)^{S_c}}{\prod_{k}p_k}\frac{1}{1+\frac{\sum_{k}p_k}{\lambda}}\!\right)\!.
\end{align}

We note that when $K=1$, i.e., single-cache system, the user freshness in (\ref{F_u_i}) reduces to the expression in \eqref{F_u(i)final}. When $K=2$, i.e., two-cache system, the user freshness in (\ref{F_u_i}) reduces to the expression in (\ref{Fu_2files}) at the top of the next page. Interestingly,  comparing  (\ref{F_u(i)final}) and (\ref{Fu_2files}), we note that freshness in a two-cache system with update rates $(c_1, c_2)$ from the source to the caches and $(u_1, u_2)$ from caches to the user, yields a smaller freshness than in a single-cache system with an update rate $c=c_1+c_2$ from the source to a cache and $u=u_1+u_2$ from the cache to the user due to the negative term in (\ref{Fu_2files}).

\begin{figure*}[t]
\begin{align}
F_u(i)&=\frac{(u_1+u_2)(c_1+c_2)}{(\lambda+u_1+u_2)(\lambda+c_1+c_2)} -\frac{\lambda}{(\lambda+u_1+u_2)(\lambda+c_1+c_2)}\left(\frac{u_2c_1}{(\lambda+u_1+c_2)}+\frac{u_1c_2}{(\lambda+u_2+c_1)}\right) \label{Fu_2files}\\
F_u(i)&=\frac{4\bar{c}\bar{u}}{(\lambda+2\bar{c} )(\lambda + 2\bar{u})}-\frac{\lambda}{(\lambda+2\bar{c} )(\lambda + 2\bar{u})}\left(\frac{(\bar{u} - b)(\bar{c} + a)}{( \bar{c} + \lambda + \bar{u} +b - a )}+\frac{(\bar{u} + b)(\bar{c} - a)}{( \bar{c} + \lambda + \bar{u} + a - b)}\right) \label{F_u2cache_ab}
\end{align}
\rule{18.1cm}{0.2mm}
\vspace{-0.8cm}
\end{figure*}

\section{Structure of the Optimal Policy } \label{sect:struc-opt-pol}
In this section, we find the optimum update rate allocation structure for general $K$ and $N$. First, we consider the system with $K=2$ caches and $N=2$ files. We denote route $k$ as the file update path from source through cache $k$ to the user. Again dropping file index $i$, let user update rates for file $i$ be $u_1$ and $u_2$ in route $1$ and route $2$, respectively, also let cache update rates in route $1$ and route $2$ be $c_1$ and $c_2$, respectively. We define the average variables as $\bar{u}=\frac{u_1+u_2}{2}$ and $\bar{c}=\frac{c_1+c_2}{2}$, and deviation from the average as $b=\frac{u_2-u_1}{2}$ and $a=\frac{c_2-c_1}{2}$. Thus, $u_1=\bar{u}-b$, $u_2=\bar{u}+b$, $c_1=\bar{c}-a$, and $c_2=\bar{c}+a$.

In the next lemma, for given user rates $u_1$ and $u_2$ (therefore, given $\bar{u}$ and $b$), and the total cache rate $2\bar{c}$, we find the optimal distribution of cache rates to maximize the freshness at the user, that is, we find the optimal $a$, $a^*$, in terms of $b$, $\bar{u}$ and $\bar{c}$.

\begin{lemma}\label{finding_optimal_a}
	In a cache update system with $K=2$ parallel caches and $N=2$ files, for given user rates $u_1$ and $u_2$, and the total cache rate $2\bar{c}$, the optimal cache rates are equal to $c_1^* = \bar{c}-a^*$ and $c_2^* = \bar{c}+a^*$ where 
	\begin{align} \label{a_optima}
    a^*= \min \bigg\{&b+\frac{(\bar{c}+\lambda+\bar{u})}{b(2\bar{c} + \lambda)} \bigg( \bar{u}(2\bar{c}+\lambda+ \bar{u})-b^2 \nonumber\\
    &-\sqrt{( \bar{u}^2-b^2)((2\bar{c} + \lambda + \bar{u})^2-b^2)}\bigg),\bar{c}\bigg\}.
    \end{align}
\end{lemma} 

\begin{Proof}
We prove the lemma by writing (\ref{Fu_2files}) equivalently as (\ref{F_u2cache_ab}) after inserting $\bar{u}$, $\bar{c}$, $a$ and $b$. Since $\bar{u}$ and $\bar{c}$ are fixed, the first term and pre-factor of the second term in (\ref{F_u2cache_ab}) are fixed. Taking the derivative of the term inside the parentheses with respect to $a$  yields the first part of the $\min$ in (\ref{a_optima}). As this critical point yields $\frac{\partial^2F_u(i)}{\partial a^2}<0$, we conclude that $a^*$ in (\ref{a_optima}) maximizes the freshness at the user. We note that $\frac{\partial a^*}{\partial b}\geq 0$, and thus, $a^*$ increases monotonically with $b$, till it reaches $\bar{c}$, after which $a^*$ is equal to $\bar{c}$, yielding (\ref{a_optima}). 
\end{Proof}

As an aside, we remark that in (\ref{a_optima}) we have $a^*\geq b$ as long as $a^*< \bar{c}$, that is, for a deviation $b$ of $u_1$, $u_2$ from their average $\bar{u}$, the optimal $a$ yields a bigger deviation for $c_1^*$, $c_2^*$ from their average $\bar{c}$. 

Next, we define $\tilde{F}_u(i)$ as the \emph{cache-update-rate-optimized} freshness, where for fixed $u_1$, $u_2$, we insert the optimal cache update rates $c_1^*$ and $c_2^*$ in (\ref{Fu_2files}). Note that  $\tilde{F}_u(i)$ is a function of $b$, $\bar{u}$ and $\bar{c}$. In the following lemma, we show that as $u_1$, $u_2$ get more \emph{lopsided}, i.e., as the difference $(u_2-u_1)$ increases, cache-update-rate-optimized freshness $\tilde{F}_u(i)$ increases. 

\begin{lemma}\label{finding_optimal_b}
	$\tilde{F}_u(i)$ is an increasing function of $b$. 
\end{lemma}
We prove Lemma~\ref{finding_optimal_b} by showing $\frac{d\tilde{F}_u(i)}{db}>0$. Lemma~\ref{finding_optimal_b} implies that lopsided update rates at the user increase the freshness. 

Next, for a $K=2$ cache system with $N=2$ files, we show that we should restrict at least one of the files to a single route, that is, lopside at least one of the files to an extreme. 

\begin{lemma}\label{shrinking_domain_size}
In a cache update system with $K=2$ caches and $N=2$ files, in the optimal policy, we need to restrict at least one file to a single route.
\end{lemma} 

\begin{figure}[t]
\centerline{\includegraphics[scale=0.42]{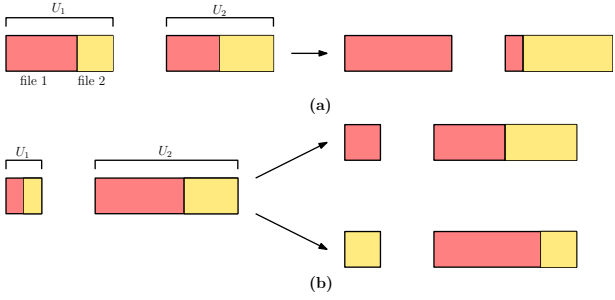}}
\vspace*{-0.1cm}
\caption{Shuffling user rates for improving freshness. (a) Freshness of both files improve (file 2 only in route 2). (b) In upper branch, freshness of file 1 decreases and of file 2 increases (file 2 only in route 2). In lower branch, freshness of file 1 increases and of file 2 decreases (file 1 only in route 2).}
\label{shuffling_user_rates}
\vspace*{-0.5cm}
\end{figure}

\begin{Proof}
Let the average rates at the caches and at the user hold values $\bar{c}_i = \frac{c_{1i}+c_{2i}}{2}$ and $\bar{u}_i = \frac{u_{1i}+u_{2i}}{2}$ for $i =1,2$ which fixes total user rates and total cache rates. Similarly, we have $u_{1i}=\bar{u}_i-b_i$ and $u_{2i}=\bar{u}_i+b_i$ which satisfies the total update rate constraints $u_{11}+u_{12}=U_1$ and $u_{21}+u_{22}=U_2$. Then, we change the update rates at the user to $u_{11}'=\bar{u}_1-b_{1}-\delta_{1}$, $u_{21}'=\bar{u}_1+b_{1}+\delta_{1}$, $u_{12}'=\bar{u}_2-b_{2}+\delta_{2}$ and $u_{22}'=\bar{u}_2+b_{2}-\delta_{2}$ such that we have $|\delta_{1}|=|\delta_{2}|$, $u_{11}'+u_{12}'=U_1$, and $u_{21}'+u_{22}'=U_2$ still hold. We analyze two cases of shuffling, shown in Fig.~\ref{shuffling_user_rates}.

In the first case, increasing $b_i$ for one file leads to increasing $b_i$ value for the other file as shown in Fig.~\ref{shuffling_user_rates}(a). As distributions of user rates for both files become lopsided simultaneously, it is a win-win situation for both files. For this case, we increase $b_i$ values of files till one file is completely in a single route. For example, in Fig.~\ref{shuffling_user_rates}(a), the user rates for the second file (shown in yellow) are $\bar{u}_2-b_{2}$ and $\bar{u}_2+b_{2}$ in route 1 and route 2, respectively. Then, we increase $b_{2}$ till $\bar{u}_2-b_{2}=0$ in route 1 and the second file is completely restricted to route 2. Such shuffling also leads to a simultaneous increase in $b_{1}$.

In the second case, increasing $b_i$ value of one file decreases $b_i$ value of the other file. This case is shown in Fig.~\ref{shuffling_user_rates}(b) where both files have larger user update rates in route $2$. In order to determine which file to prioritize, we compare
$\frac{d\tilde{F}_u(i)}{db_i}$ for both files. If $\frac{d\tilde{F}_u(1)}{db_{1}}>\frac{d\tilde{F}_u(2)}{db_{2}}$, then we prioritize improving freshness of file 1. One can show that $\frac{d^2\tilde{F}_u(i)}{db_i^2}>0$. Thus, the increase in freshness of file $1$ is always larger than the decrease in freshness of file $2$. Similarly, if $\frac{d\tilde{F}_u(2)}{db_{2}}>\frac{d\tilde{F}_u(1)}{db_{1}}$, then we increase the freshness of the second file which decreases the freshness of the first file. Thus, we need to restrict at least one file to a single route to obtain the optimum freshness.
\end{Proof}

Thus, for a $K=2$ cache and $N=2$ file system with a given set of update rates $u_{11}$, $u_{21}$, $u_{12}$ and $u_{22}$, we can shuffle these rates to increase the total freshness while keeping average rates $\bar{u}_1$, $\bar{u}_2$, $\bar{c}_1$, and $\bar{c}_2$ the same. In this process, we always end up restricting one of the files to only one route. Extending this result to a $K=2$ cache but arbitrary $N$ files case, we iteratively choose a pair of files and increase freshness of the pair by restricting one of these files to a single route. We repeat this process until we restrict $N-1$ files to a single route each. Thus, for a $K=2$ cache, arbitrary $N$ file system, only at most one file will be updated through both relays, and the remaining $N-1$ files will settle to a single relay.  

\begin{lemma}\label{lemma_merging_2_routes_K_caches}
Freshness of a file in a $K$-cache system with update rates at the cache $(c_1, c_2, c_3,\ldots,c_K)$, and at the user $(u_1, u_2, u_3, \ldots,u_K)$ is smaller than the freshness in a $(K-1)$-cache system with update rates at the cache $(c_1+c_2, c_3,\ldots,c_K)$, and at the user   
$(u_1+u_2, u_3, \ldots,u_K)$.
\end{lemma}

\begin{Proof}
With notation of Section~\ref{sect:system-model}, since freshness $F_u(i)=1-\lambda E[W_u]$, where $E[W_u]=\int_{0}^{\infty}E[W_u|T_s=t]\lambda e^{-\lambda t}dt$, we prove the lemma by showing $E[W_u^K|T_s=t]-E[W_u^{K-1}|T_s=t]\geq0$, where $K$ in $W_u^K$ denotes $K$-cache system.
\end{Proof}

Thus, given total update rates $\sum_{k=1}^{K} u_{ki} $ and $\sum_{k=1}^{K} c_{ki}$ for a file $i$, the maximum freshness is obtained by concentrating the rates in a single route to the extent possible. In the next section, we provide an approximate way of finding total update rates for files and scheduling them to individual links. 

\section{Approximate Solution} \label{sect:approx-soln}
The freshness maximization problem for our system is,
\begin{align}
\max_{c_{ki},u_{ki}} ~ &
\sum_{i=1}^{N}F_u(i) \nonumber\\
\textrm{s.t.} ~ & \sum_{k=1}^{K}\sum_{i=1}^{N}c_{ki}\leq C \nonumber\\
&\sum_{i=1}^{N}u_{ki}\leq U_k,\quad k=1,\ldots,K, \nonumber\\
&c_{ki}\geq0,\: u_{ki}\geq0,\: k=1,\ldots,K,\: i=1,\ldots,N. \label{optimization_problem}
\end{align}
This parallel-cache problem is significantly more complex than the cascade-cache problem  in \cite{bastopcu2020LineNetwork}. A Lagrangian approach as in \cite{bastopcu2020LineNetwork} seems prohibitive as it results in highly nonlinear KKT conditions. We pursue an approximate solution approach utilizing the properties of the optimal solution found in Section~\ref{shrinking_domain_size}.  

First, we construct a single-cache problem by bringing all relays together, where the source-to-cache total update constraint is $C$ and the relay-to-user total update constraint is $U=\sum_{k=1}^{K}U_k$. The optimal solution of this single-cache problem forms an upper bound for the optimum solution of our multi-cache problem, as it allows distributed relays to share update rate capacities. We denote this upper bound by $F_{ub}$.  

Second, we extract a feasible solution for our multi-cache problem from the optimum solution of the constructed single-cache problem.  We know from Lemma~\ref{lemma_merging_2_routes_K_caches} that files need to be restricted to single routes for maximum freshness. Thus, our approximate solution takes the optimum solution of the constructed single-cache problem, and distributes the update rates in the multi-cache setting in such a way that each file is updated only through a single relay to the extent possible. Let the solution of the single-cache problem be $u_i$ which is $u_i=\sum_{k=1}^{K} u_{ki}$. We assign the files in order of decreasing $u_{i}$ to one of the routes. We start with the first route and fit fully as many files as possible, till we reach a file which will not fit completely and we make it split rates with the last route (route $K$). We follow this for $K-1$ routes. If a file rate $u_i$ exceeds route capacity $U_k$, we first fill maximal full routes with it, then try to fit the remaining rate fully in the remaining routes. This leaves us with at most $K-1$ files that split rates between two routes. The remaining files go to route $K$. This approximate solution gives us a sub-optimal freshness $F_{so}$. Denoting the optimal freshness in our problem in (\ref{optimization_problem}) as $F^*$, we have
\begin{align}
F_{so}<F^*<F_{ub}
\end{align}
which means $F^*-F_{so}\leq F_{ub}-F_{so}$, i.e., the gap between the sub-optimal solution and the optimal solution is bounded by the gap between the upper bound and the sub-optimal solution. 

Next, we bound $F_{ub}-F_{so}$. We note that, in the sub-optimal policy, we assign at most $K-1$ files to two routes. From Lemma~\ref{finding_optimal_b}, freshness for file $i$ increases when $b_i$ increases, with minimum at $b_i=0$ ($a^*=0$) and maximum at $b_i=\bar{u}_i$ ($a^*=\bar{c}$). Hence, using \eqref{F_u2cache_ab}, we find an upper bound on maximum freshness loss ratio $\rho$ possible for a file due to splitting, 
\begin{align}
  \rho=\frac{F_u(i)|_{(b_i,a_i^*)=(\bar{u}_i,\bar{c}_i)}-F_u(i)|_{(b_i,a_i^*)=(0,0)}}{F_u(i)|_{(b_i,a_i^*)=(\bar{u}_i,\bar{c}_i)}},\label{freshness_loss_ratio}
\end{align}
which is equal to $\rho =\frac{\lambda_i}{2(\lambda_i+\bar{u}_i+\bar{c}_i)}<0.5$. Since $F_{ub}(i)<1$, the optimality gap is $F^*-F_{so}\leq \rho(K-1) F_{ub}(i)<0.5(K-1)$.
These $K-1$ files have low $u_i$s, owing to very high or very low $\lambda_i$s, as observed in \cite{bastopcu2020LineNetwork}. In the former case, $F_{ub}(i)$ is low, while in the latter case, $\rho$ is very low.

\section{Numerical Results} \label{sect:num-res}
We choose number of routes $K=5$, number of files $N=30$, total update rate at source $C=50$ and at caches $U=100$ where each route has  $U_k=20$. We use update arrival rates $ \lambda_i=bq^i$ at the source for $i =1,\ldots, N$, where $b>0$, $q=0.7$, and $\sum_{i=1}^{N}\lambda_i=a$, with $a=100$. Note that the update arrival rates at the source $\lambda_i$ decrease with the file index since $q<1$. 

\begin{figure}[t]
	\begin{center}
		\subfigure[]{%
			\includegraphics[width=0.49\linewidth]{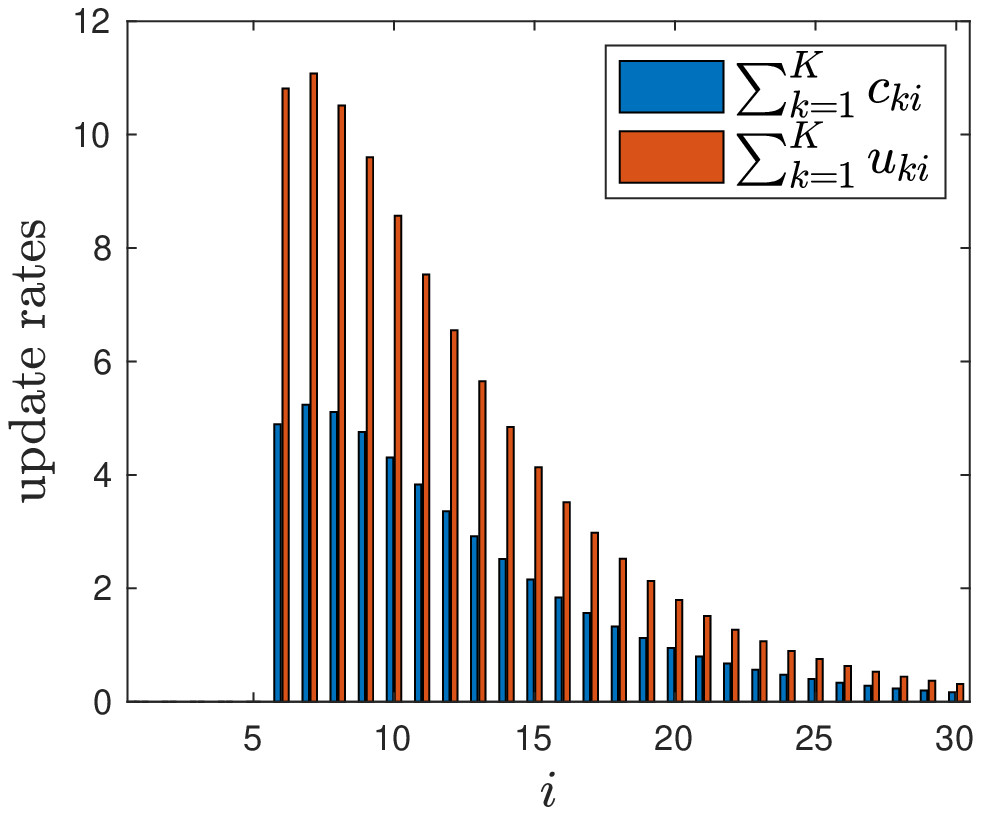}}
		\subfigure[]{%
			\includegraphics[width=0.49\linewidth]{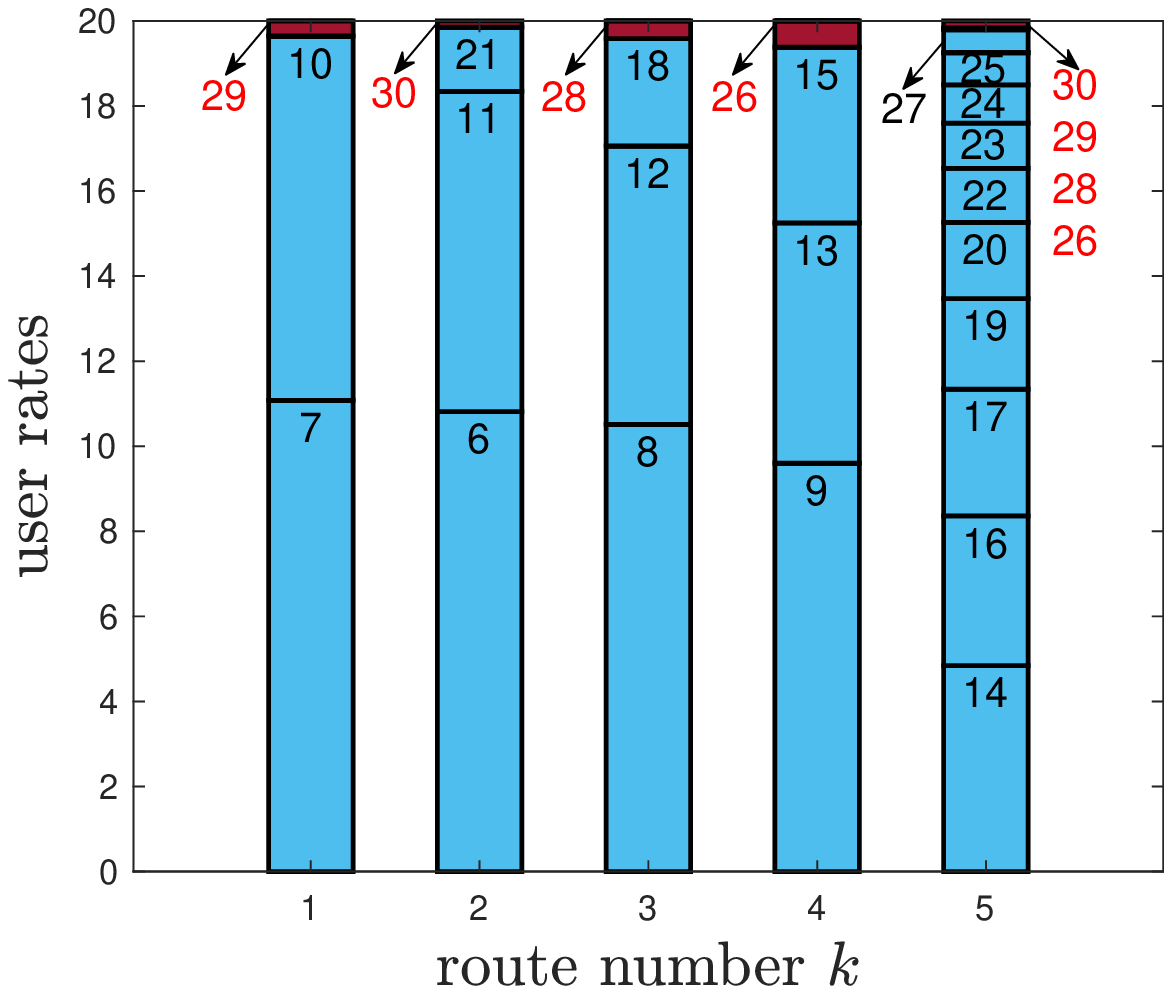}}\\ \vspace*{-0.3cm}
		\subfigure[]{%
			\includegraphics[width=0.49\linewidth]{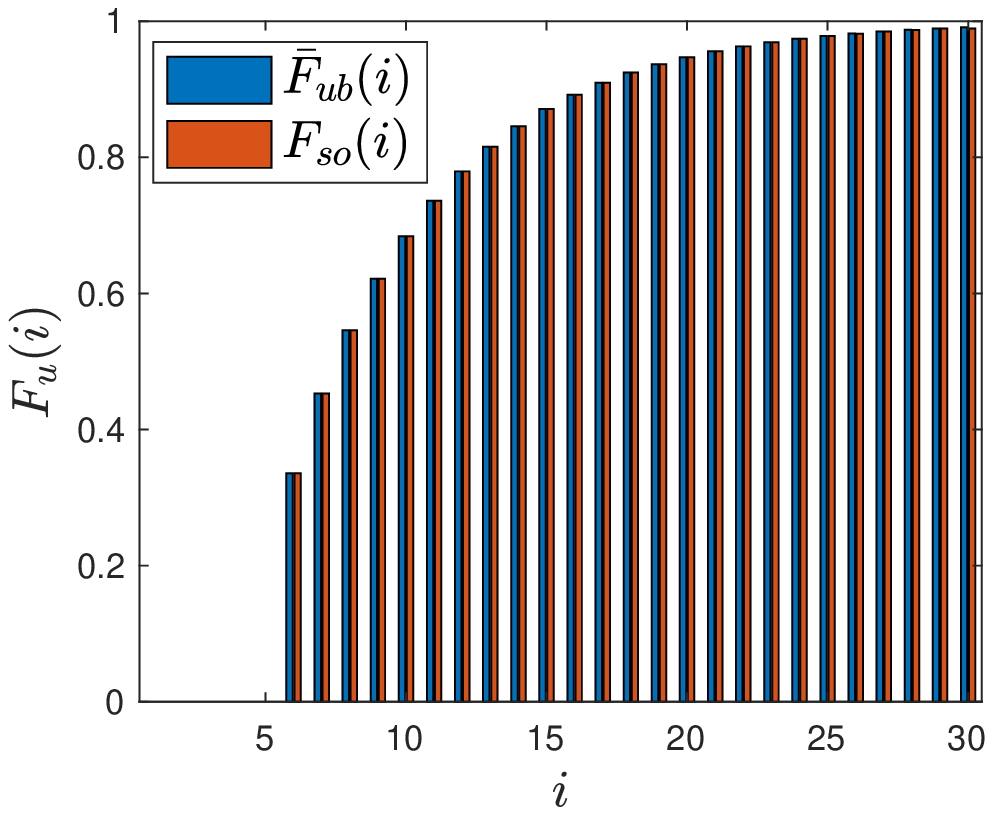}}
	\end{center}
	\vspace{-0.4cm}
	\caption{ (a) Total user rates and total cache rates obtained from the auxiliary solution, (b) route allocations for files (files with blue rates in single routes), and (c) freshness obtained for the single-cache and parallel cache systems.}
	\label{fig:simulation_30files_5routes}
	\vspace{-0.4cm}
\end{figure}

We apply alternating maximization approach described in \cite{bastopcu2020LineNetwork} to solve the auxiliary single-cache problem to obtain total update rate for file $i$ at the user $\sum_{k=1}^{K} u_{ki}$ and at the caches $\sum_{k=1}^{K} c_{ki}$ as shown in Fig.~\ref{fig:simulation_30files_5routes}(a). The total cache update rate constraint, i.e., $\sum_{k=1}^{K}\sum_{i=1}^{N}c_{ki}\leq C$, is already satisfied by both problems. In a parallel cache system, each route has its own total update rate constraint $\sum_{i=1}^{N}u_{ki}\leq U_k$, whereas  the single-cache system has only one total update rate constraint for the user, i.e., $\sum_{k=1}^{K}\sum_{i=1}^{N}u_{ki}\leq U$ where $U=\sum_{k=1}^{K}U_k$. Thus, we need to choose the user rate allocation for all files in each route as described in Section~\ref{sect:approx-soln}, with corresponding cache rates found by (\ref{a_optima}) which are shown in Fig.~\ref{fig:simulation_30files_5routes}(b). We denote the freshness at the user for the single-cache system obtained by the method in \cite{bastopcu2020LineNetwork} as $\bar{F}_{ub}$. We plot $\bar{F}_{ub}$ and $F_{so}$ in Fig.~\ref{fig:simulation_30files_5routes}(c). Even though we split the update rates among different routes for $K-1=4$ files that have some of the highest freshness (files 26, 28, 29, 30), their freshness loss is negligible as shown in Fig.~\ref{fig:simulation_30files_5routes}(c). In this system, the total freshness loss, i.e., $\bar{F}_{ub}-F_{so}$, is equal to $0.0026$, which is much smaller than the theoretical upper bound $0.5(K-1)=2$.

\bibliographystyle{unsrt}
\bibliography{ref_priyanka}

\end{document}